\documentclass[12pt]{article}
\usepackage{tcolorbox}
\usepackage{amssymb,amsmath,bm}
\usepackage[colorlinks=true, urlcolor=blue,linkcolor=blue, citecolor=blue]{hyperref}
\usepackage{mathrsfs,verbatim}
\usepackage{caption,cleveref }
\usepackage{graphicx, float}
\usepackage{constants,color,appendix,  mathtools, tikz, xcolor}
\usepackage{ulem}
\usepackage{enumerate}
\usepackage{bbm}
\usepackage{tkz-euclide}
\usepackage{tikz}
\usetikzlibrary{trees}
\usepackage{pdfpages}
\usepackage{xcolor}
\usepackage{relsize} 
\graphicspath{{./images/}}

\usetikzlibrary{calc}
\usepackage{lipsum}

\newcommand{\R}{\mathbb{R}}
\newcommand{\E}{\mathbb{E}}
\newcommand{\Pro}{\mathbb{P}}

\usepackage{appendix}
\begin{document}
\newcommand{\bea}{\begin{eqnarray}}
\newcommand{\ena}{\end{eqnarray}}
\newcommand{\beas}{\begin{eqnarray*}}
\newcommand{\enas}{\end{eqnarray*}}
\newcommand{\beq}{\begin{equation}}
\newcommand{\enq}{\end{equation}}
\def\qed{\hfill \mbox{\rule{0.5em}{0.5em}}}
\newcommand{\bbox}{\hfill $\Box$}
\newcommand{\ignore}[1]{}
\newcommand{\ignorex}[1]{#1}
\newcommand{\wtilde}[1]{\widetilde{#1}}
\newcommand{\mq}[1]{\mbox{#1}\quad}
\newcommand{\bs}[1]{\boldsymbol{#1}}
\newcommand{\qmq}[1]{\quad\mbox{#1}\quad}
\newcommand{\qm}[1]{\quad\mbox{#1}}
\newcommand{\nn}{\nonumber}
\newcommand{\Bvert}{\left\vert\vphantom{\frac{1}{1}}\right.}
\newcommand{\To}{\rightarrow}
\newcommand{\supp}{\mbox{supp}}
\newcommand{\law}{{\cal L}}
\newcommand{\Z}{\mathbb{Z}}
\newcommand{\mc}{\mathcal}
\newcommand{\mbf}{\mathbf}
\newcommand{\tbf}{\textbf}
\newcommand{\lp}{\left(}
\newcommand{\limm}{\lim\limits}
\newcommand{\limminf}{\liminf\limits}
\newcommand{\limmsup}{\limsup\limits}
\newcommand{\rp}{\right)}
\newcommand{\mbb}{\mathbb}
\newcommand{\rainf}{\rightarrow \infty}
\newtheorem{problem}{Problem}[section]
\newtheorem{exercise}{Exercise}[section]
\newtheorem{theorem}{Theorem}[section]
\newtheorem{corollary}{Corollary}[section]
\newtheorem{conjecture}{Conjecture}[section]
\newtheorem{proposition}{Proposition}[section]
\newtheorem{lemma}{Lemma}[section]
\newtheorem{definition}{Definition}[section]
\newtheorem{example}{Example}[section]
\newtheorem{remark}{Remark}[section]
\newtheorem{solution}{Solution}[section]
\newtheorem{case}{Case}[section]
\newtheorem{condition}{Condition}[section]
\newtheorem{defn}{Definition}[section]
\newtheorem{eg}{Example}[section]
\newtheorem{thm}{Theorem}[section]
\newtheorem{lem}{Lemma}[section]
\newtheorem{soln}{Solution}[section]
\newtheorem{propn}{Proposition}[section]
\newtheorem{ex}{Exercise}[section]
\newtheorem{conj}{Conjecture}[section]
\newtheorem{pb}{Problem}[section]
\newtheorem{cor}{Corollary}[section]
\newtheorem{rmk}{Remark}[section]
\newtheorem{note}{Note}[section]
\newtheorem{notes}{Notes}[section]
\newtheorem{readingex}{Reading exercise}[section]
\newcommand{\pf}{\noindent {\bf Proof:} }
\newcommand{\proof}{\noindent {\it Proof:} }
\frenchspacing


\newcommand{\mlarge}[1]{\mathlarger{\mathlarger{#1}}} 
\DeclarePairedDelimiter\abs{\lvert}{\rvert}
\DeclarePairedDelimiter\norm{\lVert}{\rVert}
\DeclarePairedDelimiter\set{\{}{\}}

\tikzstyle{level 1}=[level distance=2.75cm, sibling distance=5.65cm]
\tikzstyle{level 2}=[level distance=3cm, sibling distance=2.75cm]
\tikzstyle{level 3}=[level distance=3.9cm, sibling distance=1.5cm]

\tikzstyle{bag} = [text width=10em, text centered] 
\tikzstyle{end} = [circle, minimum width=3pt,fill, inner sep=0pt]


\title{\bf On a Notion of Outliers Based on Ratios of Order Statistics}
\author{Ahmet Zahid Balc{\i}o\u{g}lu\footnote{Y{\i}ld{\i}z Technical University, Department of Statistics, Istanbul, Turkey, e-mail: zahid.balcioglu@yildiz.edu.tr} \hspace{0.2in} O\u{g}uz G\"{u}rerk\footnote{Bo\u{g}azi\c{c}i University, Department of Mathematics, Istanbul, Turkey, e-mail:  oguz.gurerk@boun.edu.tr}} \vspace{0.25in}

\maketitle

 
\begin{abstract}

There are a number of mathematical formalisms of the term "outlier" in statistics, though there is no consensus on what the right notion ought to be. Accordingly, we try to give a consistent and robust definition for a specific type of outliers defined via order statistics. Our approach is based on ratios of partial sums of order statistics to investigate the tail behaviors of hypothetical and empirical distributions. We simulate our statistic on a set of distributions to mark potential outliers and use an algorithm to automatically select a cut-off point without the need of any further a priori assumption. Finally, we show the efficacy of our statistic by a simulation study on distinguishing two Pareto tails outside of the L\'{e}vy stable region.  

\bigskip


\bigskip


\end{abstract}

\section{Introduction}

The problem of existence of outliers\footnote{[which] are also referred to as abnormalities, discordants, deviants, or anomalies in the data mining and statistics literature \cite{Aggarwal_2015}.} or outlier detection ``[have] been recognized for a very long time, certainly since the middle of the eighteenth century. Daniel Bernoulli, writing in 1777 about the combination of astronomical observations, said:

\vspace{5px}

Is it right to hold that the several observations are of the same weight or moment, or equally prone to any and every error? $\ldots$ Is there everywhere the same probability? Such an assertion would be quite absurd, which is undoubtedly the reason why astronomers prefer to reject completely observations which they judge to be too wide of the truth, while retaining the rest and, indeed, assigning to them the same reliability. $\ldots$ I see no way of drawing a dividing line between those that are to be utterly rejected and those that are to be wholly retained; it may even happen that the rejected observation is the one that would have supplied the best correction to the others. Nevertheless, I do not condemn in every case the principle of rejecting one or other of the observations, indeed I approve it, whenever in the course of observation an accident occurs which in itself raises an immediate scruple in the mind of the observer, before he has considered the event and compared it with the other observations. If there is no such reason for dissatisfaction I think each and every observation should be admitted whatever its quality, as long as the observer is conscious that he has taken every care." \cite{Barnett_1978}.

\vspace{5px}

We refer the reader to \cite{Aggarwal_2015} for a detailed conceptual account as the amount of literature on outliers is vast. However, for an inclusive definition of an outlier, we start with a general one given by Grubbs in 1969, ``an outlying observation, or `outlier', may be merely an extreme manifestation of the random variability inherent in the data. ... On the other hand, an outlying observation may be the result of gross deviation from prescribed experimental procedure or an error in calculating or recording the numerical value"\cite{Grubbs_1969}. Hawkins in 1980 defined the concept of an outlier as ``[a]n outlier is an observation which deviates so much from the other observations as to arouse suspicions that it was generated by a different mechanism"\cite{Hawkins_1980}.
 
Now, before we start making clear what \textit{we} imply by an outlier -- an observation (or a subset of observations) in a set of data `which deviates so much from the remaining data as to arouse suspicions', we admit that ``it is a matter of subjective judgement on the part of the observer whether or not he picks out some observation (or set of observations) for scrutiny"\cite{Barnett_1978}; however, our interest and main lines of inquiry rest in identifying observations which can be characterized as \textit{extreme} in some way. In this paper, therefore, our purpose is to propose and investigate definitions of types of outliers so that we can minimize the number of data/sample-specific parameters in order for the working definition to have qualities that we expect from a mathematical definition to possess as well as to see if (and how) the new notion of outliers relates to some important results in probability theory and statistics; e.g. to the law of large numbers and the extreme value theory.

We first briefly discuss the work of Klebanov et al in \cite{Klebanov_2016} as they recently introduced and analyzed a new formulation for the notion of outliers based on order statistics due to certain drawbacks of the classical (and inherently conceptual) definitions of outliers given in \cite{Grubbs_1969}, \cite{Barnett_1978}, \cite{Hawkins_1980} by letting $X_1,X_2,\ldots,X_n$ be i.i.d. non-negative continuous random variables and denoting the corresponding order statistics by $X_{(1)}\leq X_{(2)}\leq \ldots \leq X_{(n)}$. Then $X_{(n)}$ is said to be an outlier of order $1/\kappa$ if $X_{(n -1)} \leq \kappa X_{(n)}$, where $\kappa \in (0,1)$ is some fixed number depending on the choice of the practitioner. Various properties of this new definition were investigated in their paper \cite{Klebanov_2016}. This new statistic, although giving a different look at the problem, has certain deficiencies. 

For example, it applies to only to cases with a single outlier, it is not robust in the sense that a small change in only a few points in the data may make an outlier a standard outcome, and vice versa. We generalizes this definition by considering the entire sample and looking at the ratios of partial sums of order statistics. Under the assumption that outliers are a subset of sample maxima, we consider the ratios of the form: \begin{equation}
\label{eq:intro}
\frac{\sum_{i=1}^m X_{(i)}}{\sum_{i=m+1}^n X_{(i)}}
\end{equation}
for $m \in \set{1, \dots, n}$. We define $X_{m+1}, \dots, X_n$ as outliers when the ratio (\ref{eq:intro}) is greater than a (pre-determined) threshold $\kappa$ value.

Our contributions also include a python library focused on tail index estimation and other computational methods related to heavy-tailed distributions \cite{Markovich_2008} as well as the simulations of our statistic. In \cref{section:def} we formally introduce our statistic. \Cref{section:prelim} comprises some preliminary lemmas for our calculations. In \cref{sec:calculation}, we derive the distribution function for our statistic. In \cref{section:approx} we give a concentration for the error margin in our calculations. In \cref{section:experiments}, we discuss outlier generating models and generate comparative simulations of our proposed statistic with well-known distributions. Finally, we propose an algorithmic method for the selection of the $\kappa$-threshold value and use it for distinguishing between two Pareto tails.

\section{A new notion of outliers}
\label{section:def}
Let $X_1,X_2,\ldots,X_n$ be i.i.d. random variables with cumulative distribution function $F$. We may further assume that these are absolutely continuous, and call the common p.d.f. $f$. Let $X_{(1)}, \ldots, X_{(n)}$ be the corresponding (increasing) order statistics. We are to investigate the problem of outliers in a way that the number of $\kappa$-outliers is defined via moving averages as  

\begin{equation}
\label{definition}
    \mc{O}_n \coloneqq n - \min\left\{i: \frac{1}{i}  \sum_{j=1}^i |X_{(j)}| < \kappa \frac{1}{n -i}  \sum_{j=i+1}^n |X_{(j)}|  \right\}
\end{equation}

The definition (\ref{definition}) may also be potentially useful for analyzing time series in a nonparametric way (i.e., without the normality or a similar distributional assumption). Assume that the $X_i$'s are nonnegative. We start by defining the following statistics 
\begin{equation}\label{definition2}
    T_{k,n}\equiv T_k= \sum_{i=n-k+1}^{n} X_{(i)},\quad 1 \leq k \leq n 
\end{equation}
denoting the sum of the top $k\in\{1, \dots, n\}$ order statistics from a sample of size $n$ and
\begin{equation}\label{definition3}
    S_{m,n} \equiv S_m =  \sum_{i= 1}^{m} X_{(i)},\quad 1 \leq m \leq n
\end{equation}
denoting the sum of the first $m\in\{1, \dots, n\}$ order statistics from the same sample. Subsequently, putting (\ref{definition2}) and (\ref{definition3}) together, we investigate probabilities of the form  $\Pro\left(\frac{1}{m}S_m<\kappa\frac{1}{n-m}T_{n-m}\right)$ where $\kappa\in(0,1)$ is fixed. Furthermore, when $n$ and $m$ are fixed, we may redefine $\kappa\equiv \kappa(n,m)\coloneqq\frac{m}{n-m}\kappa$, for convenience. In particular, this probability admits a closed form expression which we were able to simplify to an explicit formula for certain special cases. In order to compute the probability
$$
\Pro\left(\frac{S_m}{T_{n-m}}<\kappa\right),
$$
we exploit the Markov Property (MP) that the order statistics possess in order to compute the following conditional probability (which is slightly different than the probability we are interested in).
\begin{equation}\label{definition4}
    \Pro\left(\frac{S_{m-1}}{T_{n-m}}<\kappa \mid X_{(m)}=u\right)
\end{equation}

This is because the conditional distributions of a subset of the order statistics given another subset satisfy some really structured properties including the MP. For reference, see \cite{Reiss_1989}, \cite{Nevzorov_2001}, \cite{Nagaraja_2003}, \cite{Arnold_2008}. The following three lemmas we state are instrumental in calculating the probability defined above. They are well-known and follow from direct computations, so we omit the proofs. 

\section{Preliminaries}
\label{section:prelim}
\begin{lemma} Let $X_{1}, X_{2}, \cdots, X_{n}$ be independent observations from a continuous cdf $F$ with density $f .$ Fix $1 \leq i<j \leq n$. Then, the conditional distribution of $X_{(i)}$ given $X_{(j)}=x$ is the same as the unconditional distribution of the $i$-th order statistic in a sample of size $j-1$ from a new distribution, namely the original $F$ truncated at the right at $x .$ In notation,
$$
f_{X_{(i)} \mid X_{(j)}=x}(u)=\frac{(j-1) !}{(i-1) !(j-1-i) !}\left(\frac{F(u)}{F(x)}\right)^{i-1}\left(1-\frac{F(u)}{F(x)}\right)^{j-1-i} \frac{f(u)}{F(x)}, u<x.
$$\\
\end{lemma}

\begin{lemma} Let $X_{1}, X_{2}, \cdots, X_{n}$ be independent observations from a continuous cdf $F$ with density $f$. Fix $1 \leq i<j \leq n$. Then, the conditional distribution of $X_{(j)}$ given $X_{(i)}=x$ is the same as the unconditional distribution of the $(j-i)$-th order statistic in a sample of size $n-i$ from a new distribution, namely the original $F$ truncated at the left at $x$. In notation,
$$
f_{X_{(j)} \mid X_{(i)}=x}(u)=\frac{(n-i) !}{(j-i-1) !(n-j) !}\left(\frac{F(u)-F(x)}{1-F(x)}\right)^{j-i-1}\left(\frac{1-F(u)}{1-F(x)}\right)^{n-j} \frac{f(u)}{1-F(x)}, u>x .
$$\\
\end{lemma}

\begin{lemma}[Markov Property] Let $X_{1}, X_{2}, \cdots, X_{n}$ be independent observations from a continuous cdf $F$ with density $f$. Fix $1 \leq i<j \leq n$. Then, the conditional distribution of $X_{(j)}$ given $X_{(1)}=x_{1}, X_{(2)}=x_{2}, \cdots, X_{(i)}=x_{i}$ is the same as the conditional distribution of $X_{(j)}$ given $X_{(i)}=x_{i} .$ That is, given $X_{(i)}, X_{(j)}$ is independent of $X_{(1)}, X_{(2)}, \cdots, X_{(i-1)} .$\\
\end{lemma}

\begin{corollary}
Let $X_{1}, X_{2}, \cdots, X_{n}$ be independent observations from a continuous cdf $F$ with density $f$. Then, the conditional distribution of $X_{(1)}, X_{(2)}, \cdots, X_{(n-1)}$ given $X_{(n)}=x$ is the same as the unconditional distribution of the order statistics in a sample of size $n-1$ from a new distribution, namely the original $F$ truncated at the right at $x$. In notation,
$$
f_{X_{(1)}, \ldots, X_{(n-1)} \mid X_{(n)}=x}\left(u_{1}, \cdots, u_{n-1}\right)=(n-1) ! \prod_{i=1}^{n-1} \frac{f\left(u_{i}\right)}{F(x)}, \mbox{ where } u_{1}<\cdots<u_{n-1}<x .
$$
\end{corollary}
A similar result holds for the conditional distribution of $X_{(2)}, X_{(3)}, \cdots, X_{(n)}$ given $X_{(1)}=x$

\begin{corollary}
If $F$ is absolutely continuous, the order statistics, $X_{(1)}, \ldots, X_{(n)}$, form a (discrete time) Markov chain with transition densities:
$$
\begin{array}{r}
f_{i+1 \mid i}(y \mid x)=(n-i)\left(\frac{F(y)-F(x)}{1-F(x)}\right)^{n-i-1} \frac{f(y)}{1-F(x)}, 
\text { for } y>x ;\quad i=1, \ldots, n-1 .
\end{array}
$$
\end{corollary}

An important consequence of the two corollaries is that when conditioned on the $m$-th order statistic the sums $S_{m-1}$ and $T_{n-m}$ are independent. In particular:
\begin{proposition}\label{prop:1}
Let $F$ be absolutely continuous, then for any $1<k<n,$ the random vectors
$$
\boldsymbol{X}^{(1)}=\left(X_{(1)}, \ldots, X_{(k-1)}\right) \text { and } \boldsymbol{X}^{(2)}=\left(X_{(k+1)}, \ldots, X_{(n)}\right)
$$
are conditionally independent given that $X_{(k)}=x_{k},$ that is to say
$$
\begin{array}{c}
\Pro\left(\boldsymbol{X}^{(1)} \in B_{1}, \boldsymbol{X}^{(2)} \in B_{2} \mid X_{(k)}=x_{k}\right) = \\
\quad\quad \Pro\left(\boldsymbol{X}^{(1)} \in B_{1} \mid X_{(k)}=x_{k}\right) \Pro\left(\boldsymbol{X}^{(2)} \in B_{2} \mid X_{(k)}=x_{k}\right)
\end{array}
$$
for any Borel set $B_{1} \in \mathcal{B}\left(\mathbb{R}^{k-1}\right)$ and $B_{2} \in \mathcal{B}\left(\mathbb{R}^{n-k}\right)$. Furthermore, $$\left[S_{m-1}\mid X_{(m)}=u\right]\text{ and } \left[T_{n-m}\mid X_{(m)}=u\right]\text{ are independent as well}$$
since $S_{m-1}=\sum_{i=1}^{m-1}X_{(i)} \text{ and } T_{n-m}=\sum_{i=m+1}^{n}X_{(i)} $ are linear functions of $\boldsymbol{X}^{(1)} \text{ and } \boldsymbol{X}^{(2)}$, respectively.\\
\end{proposition}

\section{Finite sample statistics}
\label{sec:calculation}
As previous, we let $X_1,X_2,\ldots,X_n$ be i.i.d. non-negative random variables with absolutely continuous distribution function $F$, and call the common p.d.f. $f$. Let $X_{(1)}, \ldots, X_{(n)}$ be the corresponding order statistics. 

\subsection{Sum of the top order statistics}
Observe, denoting the distribution function of $X_{(i)}$ by $F_i$, that
\begin{align*}
\Pro\left(T_{k}<t\right)&=\Pro\left( \sum_{i=n-k+1}^{n} X_{(i)}<t\right)\\
&=\int \Pro\left( \sum_{i=n-k+1}^{n} X_{(i)}<t \mid X_{n-k}=u\right) d F_{X_{n-k}}(u)
\end{align*}

We know that for each $n-k<j\leq n$, the conditional distribution of $X_{(j)}$ given $X_{(n-k)}=u$ is formed from an i.i.d. sample of size $k$ having the cdf $G_{u}$ given by
$$
G_{u}(t)=\left\{\begin{array}{ll}
0, & t<u \\
\frac{F(t)-F(u)}{1-F(u)}, & t \geq u .
\end{array}\right.
$$

Hence,

$$
\Pro\left(T_{k}<t\right)=\int_{0}^{t} G_{u}^{*(k)}(t)d F_{X_{n-k}}(u),
$$
where $G_{u}^{*(k)}$ is the $k$-fold convolution of $G_{u}$.\\

We note that $T_{k}$ is also related to the selection differential, which is a familiar term in the genetics literature \cite{Crow_1970}, \cite{Burrows_1972}, \cite{Falconer_1996} given by
$$
D_{k}=\frac{1}{\sigma}\left(\frac{1}{k} \sum_{j=n-k+1}^{n} X_{(j)}-\mu\right)
$$
where $\mu$ and $\sigma$ are the population mean and standard deviation, respectively, \cite{Nagaraja_1981}.
It also serves as a test statistic for outliers in samples from normal distribution, \cite{Barnett_1978},\cite{Burrows_1972}; additionally, from \cite{Nagaraja_1982}, one can obtain the asymptotic distribution of $D_{k}$ (when $X_i$'s have finite second moment) under suitable centering and scaling if $n\rightarrow \infty$ with $k$ fixed as well as if $k=[np], 0<p<1 \text { and } n \rightarrow \infty$.\\

\textbf{Sum of the first $m$ order statistics: } Following a similar argument as before, we get:

$$
\Pro\left(S_{m}<t\right)=\int_{0}^{t} H_{u}^{*(m)}(t)d F_{X_{m+1}}(u),
$$
where $ H_{u}^{*(m)}$ is the $m$-fold convolution of the df $H_u$ given by

$$
H_{u}(t)=\left\{\begin{array}{ll}
\frac{F(t)}{F(u)}, & t \leq u\\
 0, & t>u.
\end{array}\right.
$$

\subsection{The case of $\Pro\left(\frac{S_{m-1}}{T_{n-m}}<\kappa\right)$}

Now, we are ready to compute the probability (\ref{definition4}) in which we were interested in getting a somewhat "nice" expression for 
$$
\Pro\left(\frac{S_{m-1}}{T_{n-m}}<\kappa \mid X_{(m)}=u\right)
$$

Define for a fixed $m\in\{1,2,\ldots,n-1\}$,  $$\boldsymbol{X}^{(1)}=\left(X_{(1)}, \ldots, X_{(m-1)}\right) \text { and } \boldsymbol{X}^{(2)}=\left(X_{(m+1)}, \ldots, X_{(n)}\right).$$

Then, it follows from Proposition \ref{prop:1} that $\left[\boldsymbol{X}^{(1)}\mid X_{(m)}=u\right]$, $\left[\boldsymbol{X}^{(2)}\mid X_{(m)}=u\right]$ and $ \left[S_{m-1}\mid X_{(m)}=u\right]$, $\left[T_{n-m}\mid X_{(m)}=u\right]$ are independent.\\

Note that we have 
$$
f_{\boldsymbol{X}^{(1)}\mid X_{(m)}}(\boldsymbol{x}^{(1)})\equiv f_{X_{(1)}, \ldots, X_{(m-1)} \mid X_{(m)}}\left(x_{1}, \cdots, x_{m-1}\right)=(m-1) ! \prod_{i=1}^{m-1} \frac{f\left(x_{i}\right)}{F(u)},
$$
for $ x_{1}<\cdots<x_{m-1}<x_m=u$. Also,\\
$$
f_{\boldsymbol{X}^{(2)}\mid X_{(m)}}(\boldsymbol{x}^{(2)})\equiv f_{X_{(m+1)}, \ldots, X_{(n)} \mid X_{(m)}}\left(x_{m+1}, \cdots, x_{n}\right)=(n-m) ! \prod_{i=m+1}^{n} \frac{f\left(x_{i}\right)}{1-F(u)},
$$
for $u=x_m<x_{m+1}<\cdots<x_{n}$.\\

So, let $R=S_{m-1}/T_{n-m}$, and observe that\\
$$f_{S_{m-1},T_{n-m}\mid X_{(m)}}(t_1,t_2)=f_{S_{m-1}\mid X_{(m)}}(t_1)\cdot f_{T_{n-m}\mid X_{(m)}}(t_2)$$
where
$$
f_{S_{m-1}\mid X_{(m)}}(t_1)= \textit{(m-1)-fold conv. using } f_{\boldsymbol{X}^{(1)}\mid X_{(m)}} 
$$
and
$$
f_{T_{n-m}\mid X_{(m)}}(t_2)= \textit{(n-m)-fold conv. using } f_{\boldsymbol{X}^{(2)}\mid X_{(m)}} 
$$

Hence,
\begin{align*}
    \Pro\left(R<\kappa \mid X_{(m)}=u\right)&=\Pro\left(\frac{S_{m-1}}{T_{n-m}}<\kappa \mid X_{(m)}=u\right)\\
    &=\int_{0}^{\infty} f_{T_{n-m}\mid X_{(m)}}(t_2)\left(\int_{0}^{\kappa t_2}f_{S_{m-1}\mid X_{(m)}}(t_1) d t_1\right) d t_2\\
    &= \int_{0}^{\infty} f_{T_{n-m}\mid X_{(m)}}(t_2) \cdot H_{u}^{*(m-1)}(\kappa t_2) d t_2\\
    &=:F_{R\mid m}(\kappa)
\end{align*}

where $ H_{u}^{*(m-1)}$ is the $(m-1)$-fold convolution of the df $H_u$ given by
 
$$
H_{u}(t)=\left\{\begin{array}{ll}
\frac{F(t)}{F(u)}, & t \leq u\\
 0, & t>u,
\end{array}\right.
$$\\
which can be calculated explicitly using $ f_{\boldsymbol{X}^{(1)}\mid X_{(m)}},$ but it is generally hard. (However, we will calculate it when we consider specific distributions, e.g. the exponential distribution.)\\

Differentiating $F_{R\mid m}(\kappa)$ with respect to $\kappa$, we get the pdf $f_{R\mid m}(\kappa)$:
$$
f_{R\mid m}(\kappa)=\frac{d}{d \kappa}\left[\int_{0}^{\infty} f_{T_{n-m}\mid X_{(m)}}(t_2)\left(\int_{0}^{\kappa t_2} f_{S_{m-1}\mid X_{(m)}}(t_1) d t_1\right) d t_2\right]
$$

$$
f_{R\mid m}(\kappa)=\int_{0}^{\infty} f_{T_{n-m}\mid X_{(m)}}(t_2) \cdot f_{S_{m-1}\mid X_{(m)}}(\kappa t_2) t_2 d t_2
$$
Therefore,
\begin{align*}
    \Pro\left(R<\kappa\right)&=\int_{0}^{\infty} \Pro\left(R<\kappa \mid X_{(m)}=u\right)dF_m(u)\\
    &=\int_{0}^{\infty}F_{R\mid m}(\kappa)dF_m(u)\\
    &=\int_{0}^{\infty}\left(\int_{0}^{\infty} f_{T_{n-m}\mid X_{(m)}}(t_2)\left(\int_{0}^{\kappa t_2}f_{S_{m-1}\mid X_{(m)}}(t_1) d t_1\right) d t_2\right)dF_m(u)\\
    &=\int_{0}^{\infty}\left(\int_{0}^{\infty} f_{T_{n-m}\mid X_{(m)}}(t_2)\cdot H_{u}^{*(m-1)}(\kappa t_2) d t_2\right)dF_m(u)
\end{align*}
where $dF_m(u)=f_{m}(u)du=\frac{n !}{(m-1) !(n-m) !} F^{m-1}(u)[1-F(u)]^{n-m} f(u)du$.

\subsection{Application: the case of the exponential distribution}

Let the parent distribution be the std. exponential; i.e., $X_i\sim$Exp(1) for each $i=1,2,\ldots,n$. Then, it is well-known that (e.g., \cite{Nagaraja_2003}, \cite{Nevzorov_1984})
$$X_{(i)}\stackrel{d}=\sum_{k=1}^{i}\frac{Z_k}{n-k+1}\hspace{2pt},$$ $ \text{where } Z_k\sim \text{Exp(1)}\text{ for each } k\geq1. \text{ Now, define }\widehat{Z_k}\coloneqq\frac{Z_k}{n-k+1} \text{ so that }\newline \widehat{Z_k}\sim$Exp($\lambda_{k}^{-1}),\hspace{5pt} \lambda_{k}=(n-k+1).$ Hence,
\begin{align*}
    T_{n-m}=\sum_{i=m+1}^{n}X_{(i)}&\stackrel{d}=\sum_{i=m+1}^{n}\sum_{k=1}^i \widehat{Z_k}\\
    &=(n-m)\sum_{k=1}^{m+1}\widehat{Z_k}+\sum_{k=m+2}^{n}(n-k+1)\widehat{Z_k}\\
    &= \sum_{k=1}^{m+1}\beta_k Z_k + W
\end{align*}
$\quad \text{where }\hspace{2pt}\beta_k\coloneqq\frac{n-m}{n-k+1}=(n-m)\lambda_k^{-1}\hspace{2pt}\text{ and }\hspace{2pt} W\sim Gamma(n-m-1,1)$.\\

Now, consider the theorem below by Jasiulewicz, H. and Kordecki, W. to find the pdf of\hspace{1.5pt} $\sum_{k=1}^{m+1}\beta_k Z_k$ in the expansion of $T_{n-m}$.

\begin{theorem}\cite{Jasiulewicz_2003}\label{thm:Jasiulewicz_2003}
Let $X_{1}, \ldots, X_{n}$ be $n$ independent random variables such that every $X_{i}$ has a probability density function $f_{X_{i}}$ given by
$$
f_{X_{i}}(t)\coloneqq\beta_{i} \exp \left(-t \beta_{i}\right) \mathbbm{1}_{(0, \infty)}(t)
$$
for all real number $t,$ where the parameter $\beta_{i}$ is positive for all $i=$ $1,2, \ldots, n$. We suppose that the parameters $\beta_{i}$ are all distinct. Then the sum $S_{n}$ has the following probability density function:
$$
f_{S_{n}}(t)=\sum_{i=1}^{n} \frac{\beta_{1} \ldots \beta_{n}}{\prod_{j=1 \atop j \neq i}^{n}\left(\beta_{j}-\beta_{i}\right)} \exp \left(-t \beta_{i}\right) \mathbbm{1}_{(0, \infty)}(t)
$$
for all $t \in \mathbb{R}$.\\\end{theorem}

So, by Theorem \ref{thm:Jasiulewicz_2003} we find the pdf of $L\coloneqq\sum_{k=1}^{m+1}\beta_k Z_k$ in the expansion of $T_{n-m}$ by noting that the $\beta_i$'s in the theorem correspond to $\frac{\lambda_i}{n-m}$ in our notation. Therefore,

\[
f_L(x) = \left(\frac{\lambda_1}{n-m}\right)\left(\frac{\lambda_2}{n-m}\right)\cdots\left(\frac{\lambda_{m+1}}{n-m}\right)\sum_{i=1}^{m+1}\mlarge{\Psi}_{i,m+1}\cdot e^{-(\frac{\lambda_i}{n-m})x}\hspace{4pt},
\]
where 
$$
\mlarge{\Psi}_{i,m+1}^{-1}=\prod_{j=1\atop j\neq i}^{m+1}\left(\frac{\lambda_j-\lambda_i}{n-m}\right)=\frac{1}{(n-m)^m}(\lambda_1-\lambda_i)\cdots(\lambda_{i-1}-\lambda_i)(\lambda_{i+1}-\lambda_i)\cdots(\lambda_{m+1}-\lambda_i)
$$
So,
$$
f_L(x)=\frac{\lambda_1\lambda_2\cdots\lambda_{m+1}}{(n-m)^{m+1}}\cdot(n-m)^m\sum_{i=1}^{m+1}\psi_{i,m+1}\cdot e^{-(\frac{\lambda_i}{n-m})x}\hspace{4pt},
$$
where 
$$\psi_{i,m+1}^{-1}=(\lambda_1-\lambda_i)\cdots(\lambda_{i-1}-\lambda_i)(\lambda_{i+1}-\lambda_i)\cdots(\lambda_{m+1}-\lambda_i)$$\\
and $\mlarge{\Psi}=(n-m)^{m}\cdot\psi_{i,m+1}$\\

Then,
$$
f_L(x)=\frac{\lambda_1\lambda_2\cdots\lambda_{m+1}}{n-m}\sum_{i=1}^{m+1}\psi_{i,m+1}\cdot e^{-(\frac{\lambda_i}{n-m})x}
$$

Now, using convolution formula, we can compute $f_{T_{n-m}}$ explicitly: 
$$
f_{T_{n-m}}(t)=\int_0^\infty f_L(t-x)f_W(x) dx\hspace{2pt},
$$
$\textit{where }\hspace{2pt} W\sim Gamma(n-m-1,1)$.

\begin{align*} 
    f_{T_{n-m}}(t)&=\frac{\lambda_1\lambda_2\cdots\lambda_{m+1}}{n-m}\sum_{i=1}^{m+1}\psi_{i,m+1}\int_0^t e^{-\frac{\lambda_i}{n-m}(t-x)}\frac{1}{(n-m-2)!}e^{-x}x^{n-m-2}dx\\
    &=\frac{1}{(k-1)!} \frac{\lambda_1\lambda_2\cdots\lambda_{m+1}}{k+1}\sum_{i=1}^{m+1}\psi_{i,m+1}e^{-\frac{\lambda_i}{k+1}t}\int_0^t e^{\left(\frac{\lambda_i}{k+1}-1\right)x}x^{k-1}dx\hspace{2pt},
\end{align*}
where we put $n-m-1\equiv k$ (for convenience\footnote{For a given $k$, one can evaluate the integral: 
$$
\int_0^t e^{\left(\frac{\lambda_i}{k+1}-1\right)x}x^{k-1}dx= t^k \left(t-\frac{\lambda_i t}{k+1}\right)^{-k} \left(\Gamma (k)-\Gamma \left(k,t-\frac{\lambda_i t}{k+1}\right)\right)
$$}) and $t>0$.\\

Recall that we wanted to get an expression for $\Pro\left(R<\kappa\right)\text{, where }R=S_{m-1}/T_{n-m}$:
\begin{align*}
    \Pro\left(R<\kappa\right)&=\int_{0}^{\infty} \Pro\left(R<\kappa \mid X_{(m)}=u\right)dF_m(u)\\
    &=\int_{0}^{\infty}\left(\int_{0}^{\infty} f_{T_{n-m}\mid X_{(m)}}(t_2)\cdot H_{u}^{*(m-1)}(\kappa t_2) d t_2\right)dF_m(u)
\end{align*}
where $dF_m(u)=f_{m}(u)du=\frac{n !}{(m-1) !(n-m) !} (1-e^{-u})^{m-1}e^{-u(n-m)}e^{-u}du$, and $ H_{u}^{*(m-1)}$ is the $(m-1)$-fold convolution of the df $H_u$ given by
$$
H_{u}(t)=\left\{\begin{array}{ll}
\frac{1-e^{-t}}{1-e^{-u}}, & t \leq u\\
 0, & t>u.
\end{array}\right.
$$

Hence, an explicit expression for our target probability is available as all the ingredients are ready (up to scaling) to be employed.

\section{Approximating $\Pro\left(S_m/T_{n-m}\leq\kappa\right)$}
\label{section:approx}

As usual, for  $X_1,X_2,\ldots,X_n$ non-negative i.i.d. random variables with absolutely continuous distribution function $F$, letting 
$$
    T_{k,n}\equiv T_k= \sum_{i=n-k+1}^{n} X_{(i)},\quad 1 \leq k \leq n 
$$ and
$$
    S_{m,n} \equiv S_m =  \sum_{i= 1}^{m} X_{(i)},\quad 1 \leq k \leq n
$$
observe, for $m=1,2,\ldots,n-1$, that
\begin{align*}
    \frac{S_m}{T_{n-m}} &= \frac{S_{m-1}+X_{(m)}}{T_{n-m}}\\
    &= \frac{S_{m-1}}{T_{n-m}} + \frac{X_{(m)}}{T_{n-m}}
\end{align*}
where $T_{n-m} = \sum_{i=m+1}^n X_{(i)}$.\\

Note that $$\frac{X_{(m)}}{T_{n-m}} = \frac{X_{(m)}}{\sum_{i=m+1}^n X_{(i)}}\leq \frac{1}{n-m}\quad\textit{ almost surely.}$$

Thus, we have

$$\Pro\left(\frac{S_{m-1}}{T_{n-m}}\leq\kappa \right) \leq \Pro\left(\frac{S_m}{T_{n-m}}\leq\kappa\right) \leq \Pro\left(\frac{S_{m-1}}{T_{n-m}}\leq\kappa+\frac{1}{n-m}\right)$$

\vspace{20pt}

Recall that we have, for $R=S_{m-1}/T_{n-m}$, that
$$f_{S_{m-1},T_{n-m}\mid X_{(m)}}(t_1,t_2)=f_{S_{m-1}\mid X_{(m)}}(t_1)\cdot f_{T_{n-m}\mid X_{(m)}}(t_2)$$
So,
\begin{align*}
    \Pro\left(R<\kappa + \frac{1}{n-m} \mid X_{(m)}=u\right)&=\Pro\left(\frac{S_{m-1}}{T_{n-m}}<\kappa + \frac{1}{n-m} \mid X_{(m)}=u\right)\\
    &=\int_{0}^{\infty} f_{T_{n-m}\mid X_{(m)}}(t_2)\left(\int_{0}^{\left(\kappa+\frac{1}{n-m}\right) t_2}f_{S_{m-1}\mid X_{(m)}}(t_1) d t_1\right) d t_2\\
\end{align*}
Now note that
\begin{align}
    \int_{0}^{\left(\kappa+\frac{1}{n-m}\right) t_2}f_{S_{m-1}\mid X_{(m)}}(t_1) d t_1 &= \int_{0}^{\kappa t_2}f_{S_{m-1}\mid X_{(m)}}(t_1) d t_1 + \int_{\kappa t_2}^{\left(\kappa+\frac{1}{n-m}\right) t_2}f_{S_{m-1}\mid X_{(m)}}(t_1) d t_1 \nonumber\\
    &\leq\int_{0}^{\kappa t_2}f_{S_{m-1}\mid X_{(m)}}(t_1) d t_1 + \frac{t_2}{n-m} \label{eq:concentration1}
\end{align}

Also, 
\begin{equation}
\label{eq:concentration2}
\int_{0}^{\infty} f_{T_{n-m}\mid X_{(m)}}(t_2)\frac{t_2}{n-m}dt_2 = \frac{\E[T_{n-m}\mid X_{(m)}]}{n-m}
\end{equation}

Therefore,
\begin{align*}
     \Pro\left(R<\kappa + \frac{1}{n-m} \right) &= \int_0^\infty\Pro\left(R<\kappa + \frac{1}{n-m} \mid X_{(m)}=u\right)dF_m(u)\\
     &\leq \int_0^\infty\left(\Pro\left(R<\kappa \mid X_{(m)}=u\right) + \frac{\E[T_{n-m}\mid X_{(m)}]}{n-m}\right)dF_m(u) & \mbox{by (\ref{eq:concentration1}) and (\ref{eq:concentration2}}),\\
     &= \Pro\left(R<\kappa\right) + \frac{1}{n-m}\int_0^\infty \E[T_{n-m}\mid X_{(m)}=u] dF_m(u)\\
     &= \Pro\left(R<\kappa\right) + \frac{1}{n-m}\E[T_{n-m}]\\
     &= \Pro\left(R<\kappa\right) + o(1),\quad n\to\infty 
\end{align*}

since if $X_i'$s have finite moment, then so is $T_{n-m}$, so that $\frac{\E[T_{n-m}]}{n-m} = o(1), \quad n\to\infty$. So, the rough approximation above gives an error bound vanishing in the order of $1/n$.

\section{Simulations and Experiments}
\label{section:experiments}
Empirically it is important to choose a good value of $\kappa$ which can distinguish between normal and anomalous observations. There are a few points which are of concern. First for a given distribution, a good $\kappa$ value should be able to distinguish between the centre and the tail of the distribution. Informally, $\kappa$ value should be natural to choose. Furthermore, it is important that, for a given distribution, the concentration of potential $\kappa$ values should be tight, potentially depending on the family of the given distribution.

Consequently, it is of importance to know how the statistic $R = \frac{S_{m}}{T_{n-m}}$ is distributed. We expect to see similar values of $\kappa$ on lower quantiles for most distributions. The differences between the $\kappa$ values should increase gradually towards the tail. Finally, at the tail we expect the $\kappa$ values to differ the most, with concentrations dependent on the tail index. From another perspective, when considering the moving averages of our statistic for the sample, we expect to have a sharp increase towards the end of the tail.

We will show simulations of $R$ for a set of distributions and anomaly generating models. As most outlier generator models depend upon the exponential distribution, we will use the exponential distribution for comparison. For anomaly generating model we will use the identified outliers model, in which observations \(X_1, \dots, X_n\) are not i.i.d. but some $k \in \{1, \dots, n\}$ come from a separate distribution. We will consider the simplistic exponential case as described below \cite{Balakrishnan_2019}.

\paragraph{Identified Outliers Model} $X=\left\{X_i,\dots, X_{n-k} : 1-e^{-x / \theta} ; \theta>0\right\}, k$ is known and fixed, say for simplicity let's suppose $k=1$, $X_{1}, \ldots, X_{n-k}$ are independent and the index of the contaminant is also known. If we assume further that the distribution function of the contaminant is
\[
G(x)=F\left(b^{-1} x\right)=1-e^{-(b \theta)^{-1} x}, \quad x \in \R
\]
for some $b \geq 1$, then, without loss of any generality, the joint distribution function of $X_{1}, \ldots, X_{n}$ is given by
\[
F_{\left(X_{1}, \ldots, X_{n}\right)}\left(x_{1}, \ldots, x_{n}\right)
= \left[\prod_{i=1}^{n-1}\left(1-e^{-x_{i} / \theta}\right)\right]
\left(1-e^{-x_{n} / b \theta}\right)\]
for $x>0$, $\theta>0$, $b\geq1$.

The simulations for the identified outliers model. In these sets of simulations we took $k=100$, $\theta=1$ for the original sample, and $b=3$ for outliers.

\paragraph{Exponential Distribution}
The simulations for exponential distribution, $f(x;\theta)= \theta e^{-x\theta}$, with $\theta = 1$.

\paragraph{Half-Normal Distribution} We will also use the absolute value of the normal distribution, the half-normal distribution, as a way to quantify the outliers in a normally distributed sample. The behaviour of moving averages of $R$-statistic for the normal distribution is specifically relevant for any empirical study. As characterizing the tail and cut-off of the normal distribution has many applications. For the simulations of Half-Normal distribution, we will use the standard normal  $Z \sim N(0, 1)$ and simulate $Y = \abs{Z}$.

Simulations are generated using the following procedure, first we generate and sort an i.i.d. sample of size $n=1000$. Then we compute the $R$-statistic for some values of $m$ in order to  better observe the changes in the behaviour of the statistic. We choose the median, $5^{th}$ percentile, and $95^{th}$ percentile points as indicators. This procedure was repeated $10000$ times. The values obtained are given in the figures \ref{plot:distributions2} below. We also provide in figures \ref{plot:distributions} moving averages of $R$-statistic throughout a sample of $n=1000$.

\begin{figure}[H]
\begin{minipage}{\linewidth}
  \centering
  \begin{tabular}{cc}
\textbf{Exponential Distribution} &
\textbf{Identified Outliers Model} \\
\includegraphics[width=\textwidth/2]{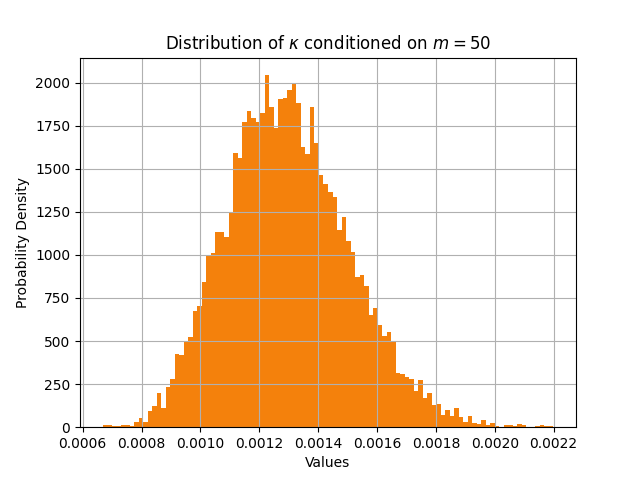}
&\includegraphics[width=\textwidth/2]{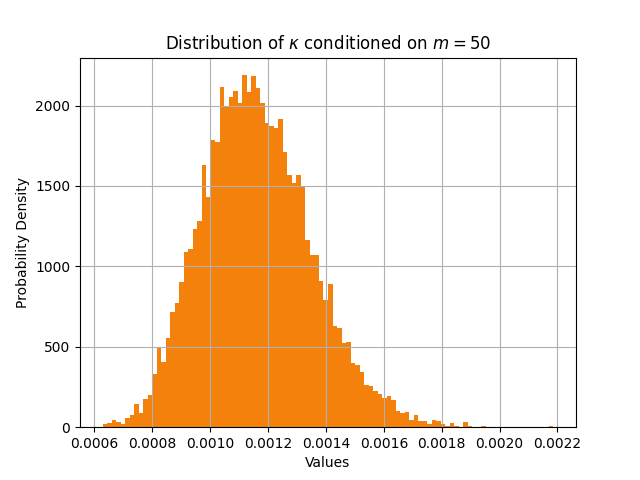} \\
\\
\includegraphics[width=\textwidth/2]{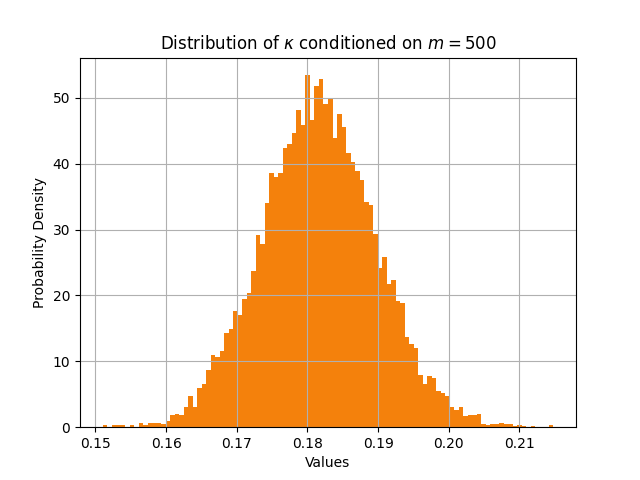}
&\includegraphics[width=\textwidth/2]{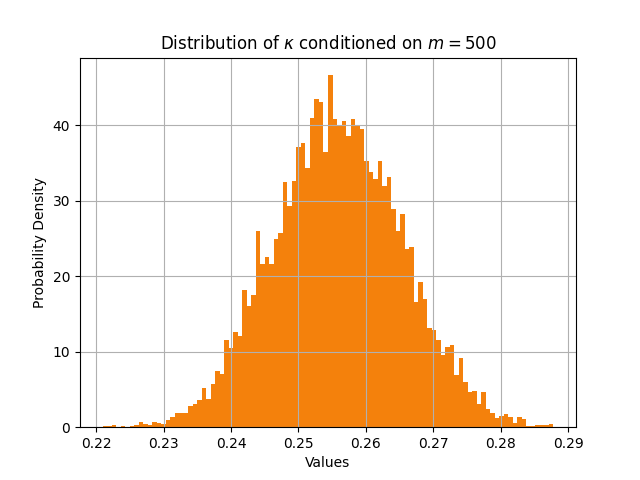} \\
\includegraphics[width=\textwidth/2]{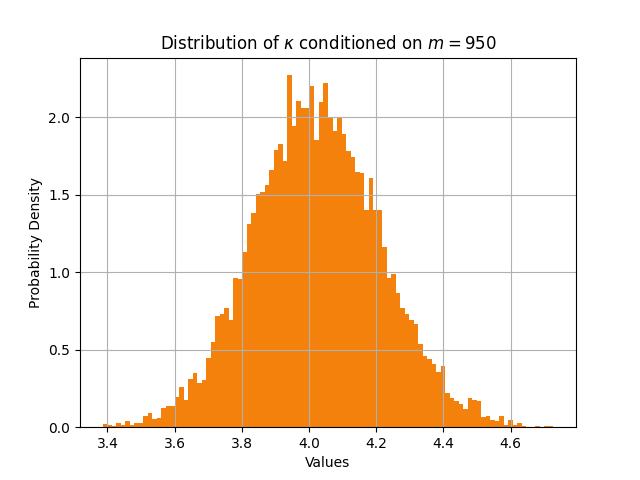}
&\includegraphics[width=\textwidth/2]{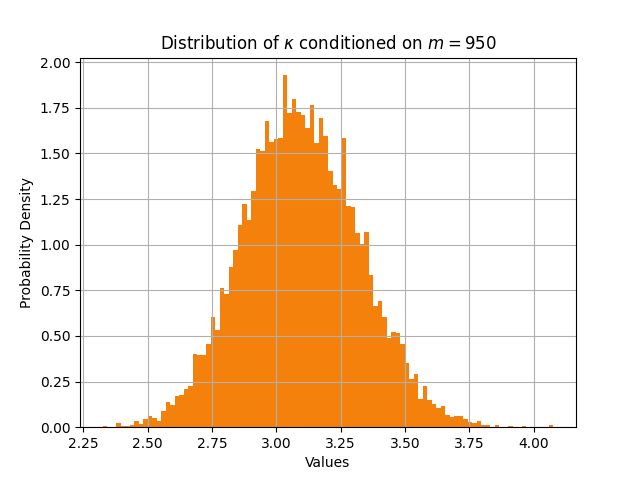} \\
  \end{tabular}
  \end{minipage}
\caption{Distribution of the $R$ statistic across 10 000 samples for fixed $m=50$, $500$, and $950$.}
\label{plot:distributions2}
\end{figure}

\begin{figure}[H]
\begin{minipage}{\linewidth}
\centering
    \begin{tabular}{ccc}
    \textbf{Half-Normal Distribution} &
    \textbf{Exponential Distribution} &
    \textbf{Identified Outliers Model} \\
    \includegraphics[width=\textwidth/3]{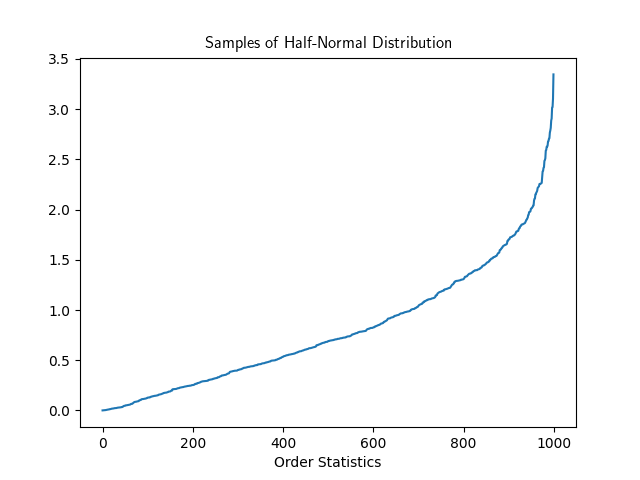}
    &\includegraphics[width=\textwidth/3]{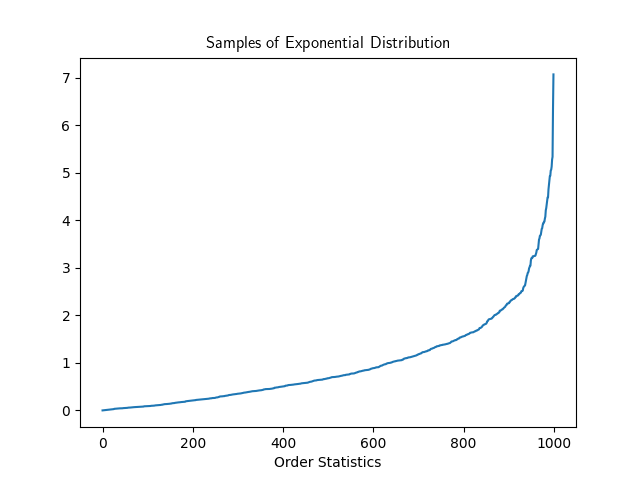}
    &\includegraphics[width=\textwidth/3]{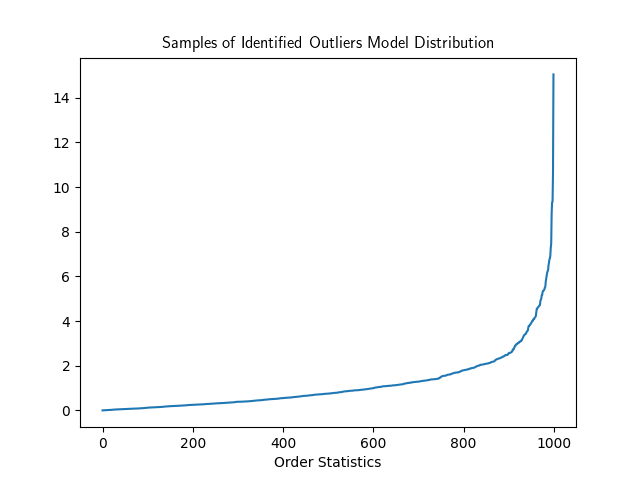} \\
    \includegraphics[width=\textwidth/3]{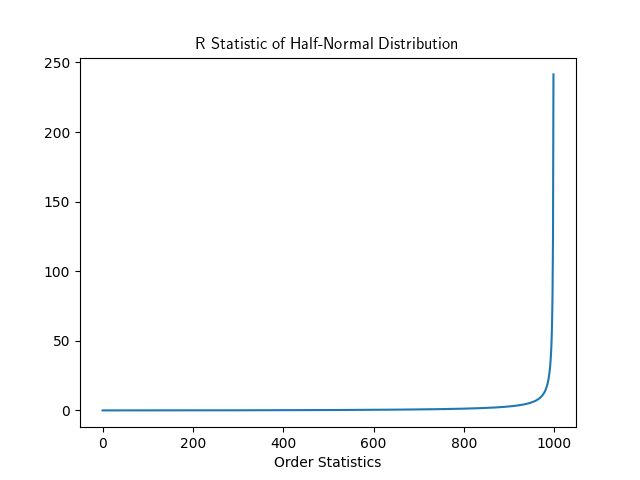}
    &\includegraphics[width=\textwidth/3]{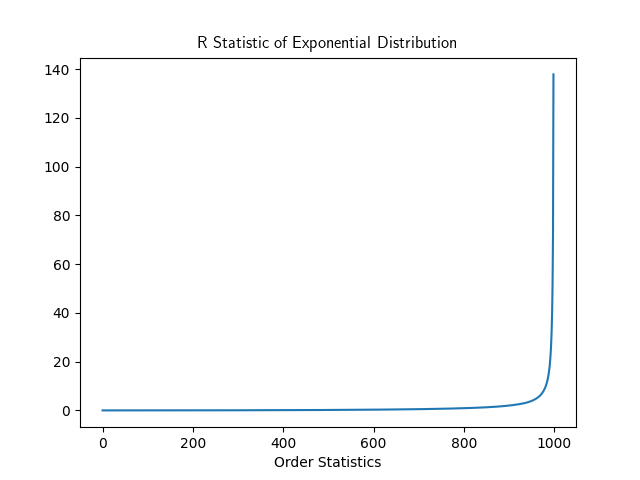}
    &\includegraphics[width=\textwidth/3]{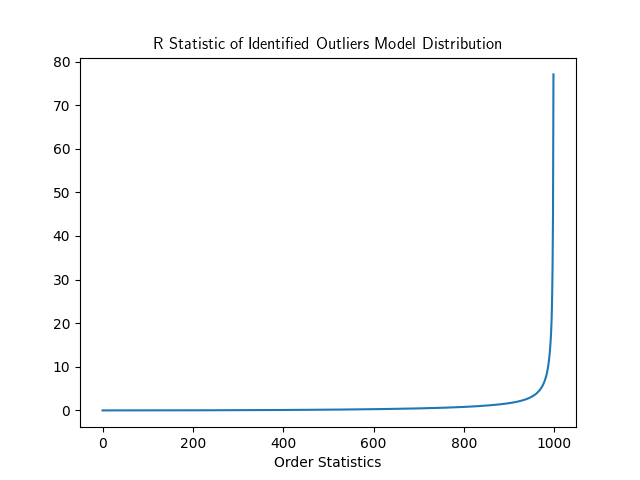} \\
    \end{tabular}
\end{minipage}
\caption{Samples of $n$=1000 and corresponding $R$ statistic from exponential distribution and identified outliers model.}
\label{plot:distributions}
\end{figure}

\subsection{Automatic selection of $\kappa$-threshold}
Ideally, we would like to define the cut-off point with respect to the derivatives of the distribution function of our statistic, which would then allow us to define a theoretical cut-off point for any sample. However, there is no established well-defined notion for an elbow point. The literature of elbow detection\footnote{Also called knee or kneecap detection.} is therefore algorithmically focused \cite{Gomes_2018}.

The elbow detection literature is based on the following pointwise definition of the curvature of a function. Most works than define the elbow as the point of maximum curvature, and use algorithmic means to calculate it.

\begin{definition}[Curvature of a function \cite{Kneedle_2011}]
\label{def:curvature}
For any continuous function $f$, there exists a standard closed-form $K_{f}(x)$ that defines the curvature of $f$ at any point as a function of its first and second derivative:
\[K_{f}(x)=\frac{f^{\prime \prime}(x)}{\left(1+f^{\prime}(x)^{2}\right)^{\frac32}}\]
\end{definition}

However, the \cref{def:curvature} of curvature does extend easily to discrete data, instead \cite{Kneedle_2011}, \cite{Tolsa_2000}, and \cite{Gomes_2018} use Menger curvature, which is defined for three points as the curvature of the circle circumscribed by those points. There are also angle-based \cite{Kneedle_2011} and exponentially weighted moving average (EWMA) based methods \cite{Albrecht_2006} which use the differences between successive points and EWMA smoothing to check deviations from arrival times respectfully.

We will use the kneedle detection algorithm for estimating the "elbow point" of our statistic \cite{Kneedle_2011}. The kneedle algorithm uses dynamic first derivative threshold, in combination with he IsoData \cite{Ridler_1978} to find the elbows of a discrete data. It can work on discrete datasets and has a sensitivity parameter which can be fine-tuned for how sensitively a knee is to be detected. While the smaller values of the sensitivity parameter respond to quick change, the larger values are more robust.

\begin{figure}[H]
\begin{minipage}{\linewidth}
\centering
    \begin{tabular}{ccc}
    \textbf{Half-Normal Distribution} & \textbf{Exponential Distribution} &
    \textbf{Identified Outliers Model} \\
    \includegraphics[width=\textwidth/3]{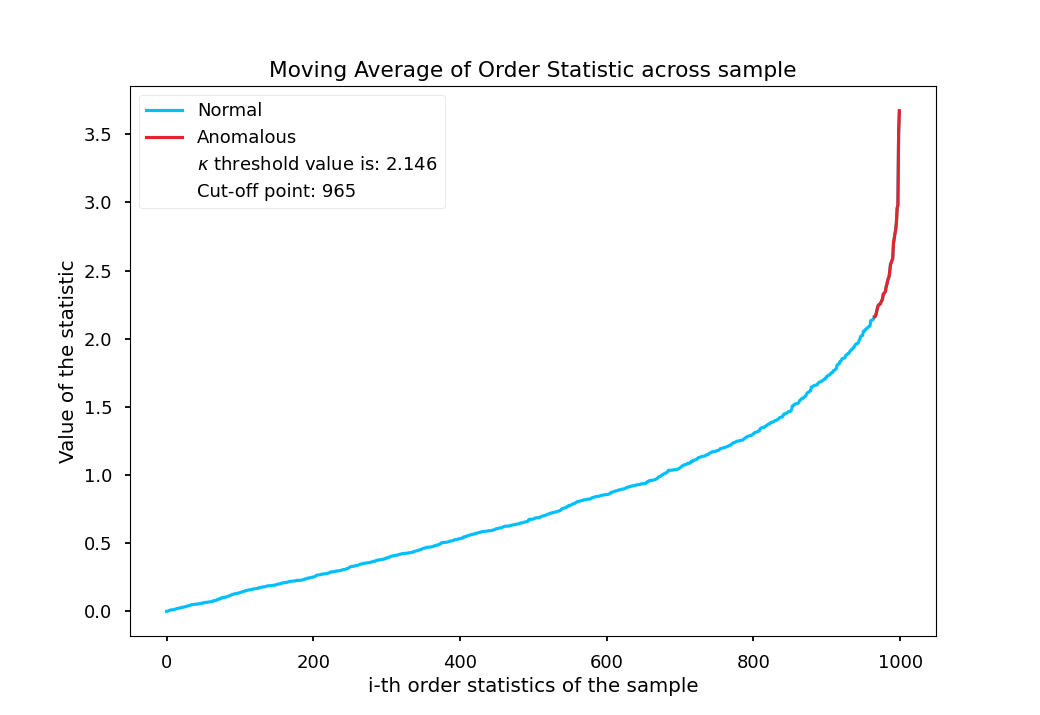}
    & \includegraphics[width=\textwidth/3]{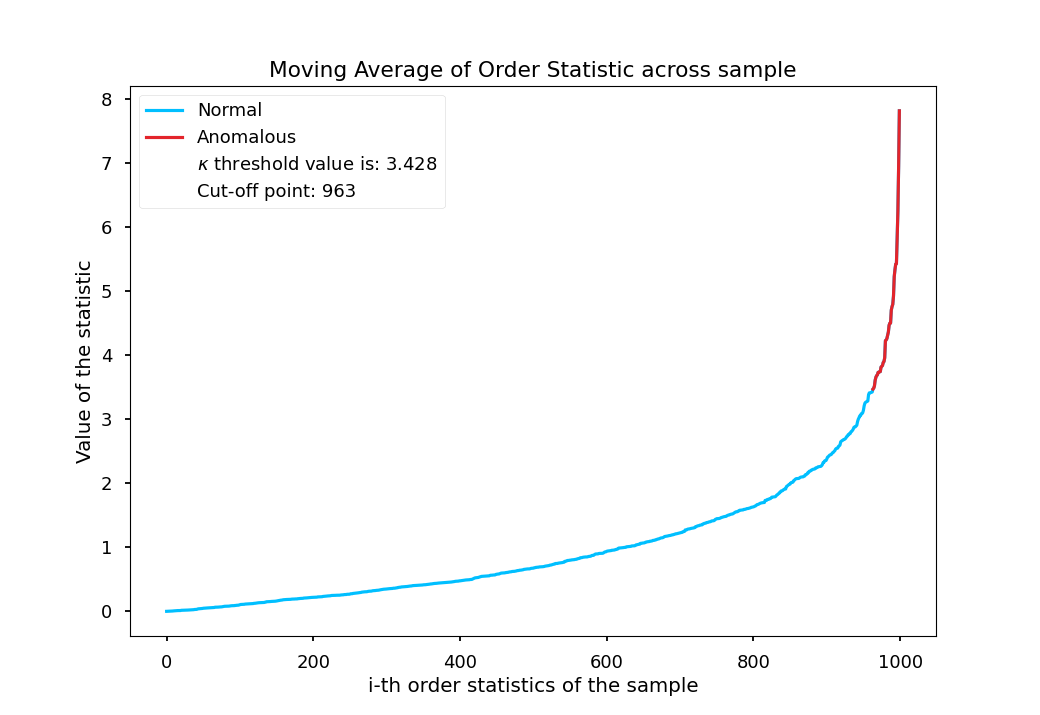}
    &\includegraphics[width=\textwidth/3]{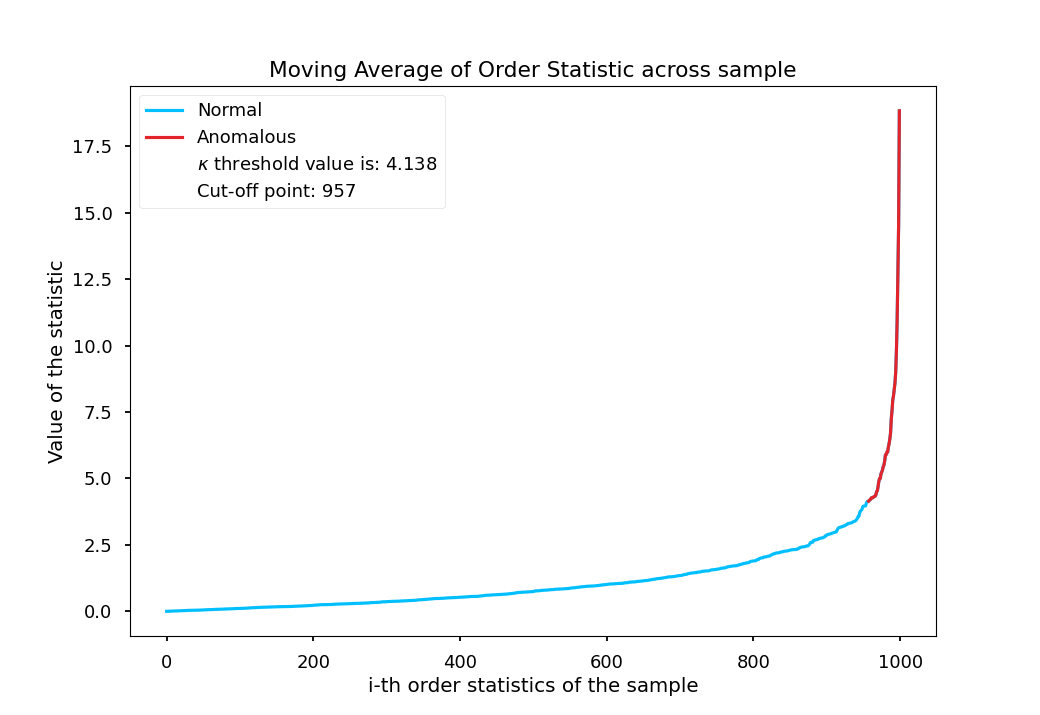} \\
    \includegraphics[width=\textwidth/3]{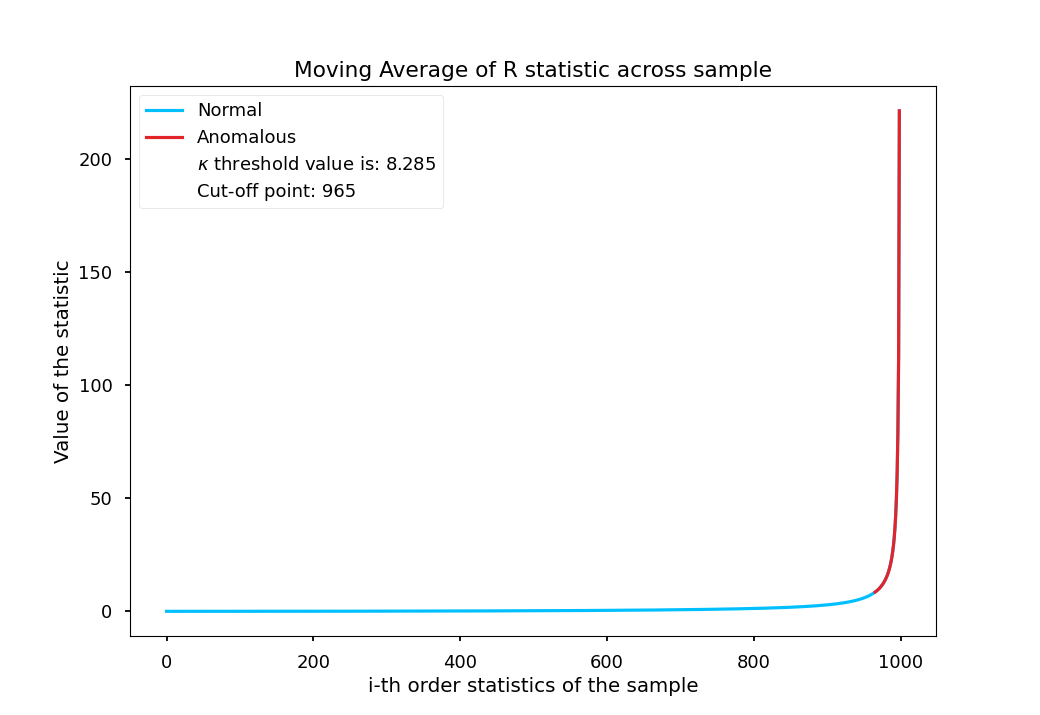}
    & \includegraphics[width=\textwidth/3]{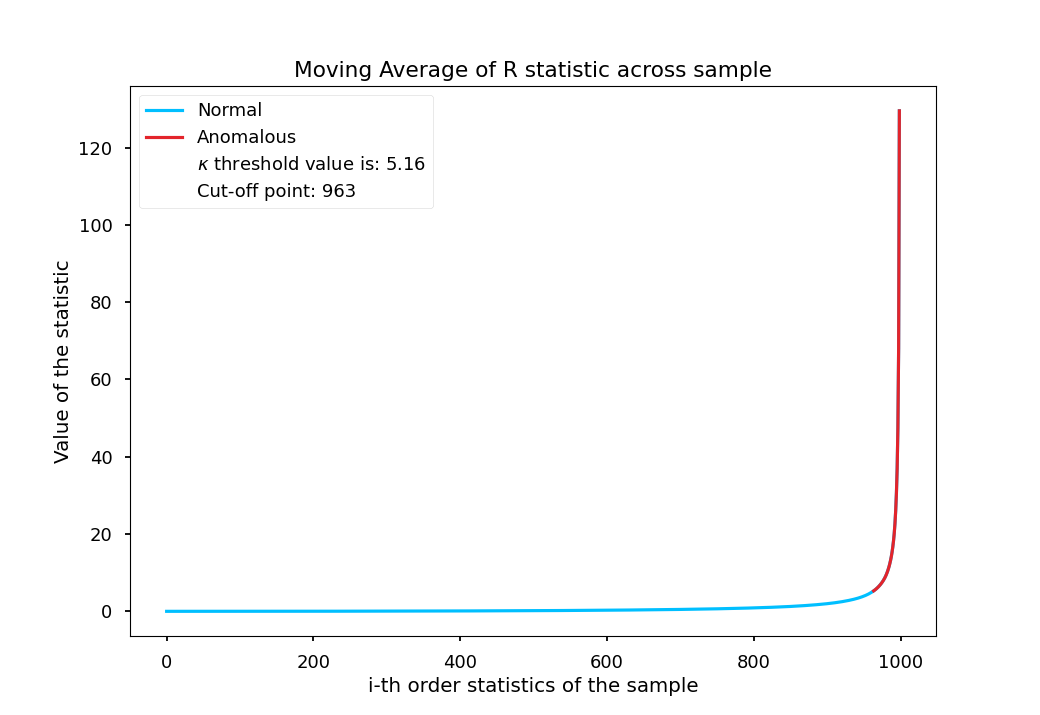}
    &\includegraphics[width=\textwidth/3]{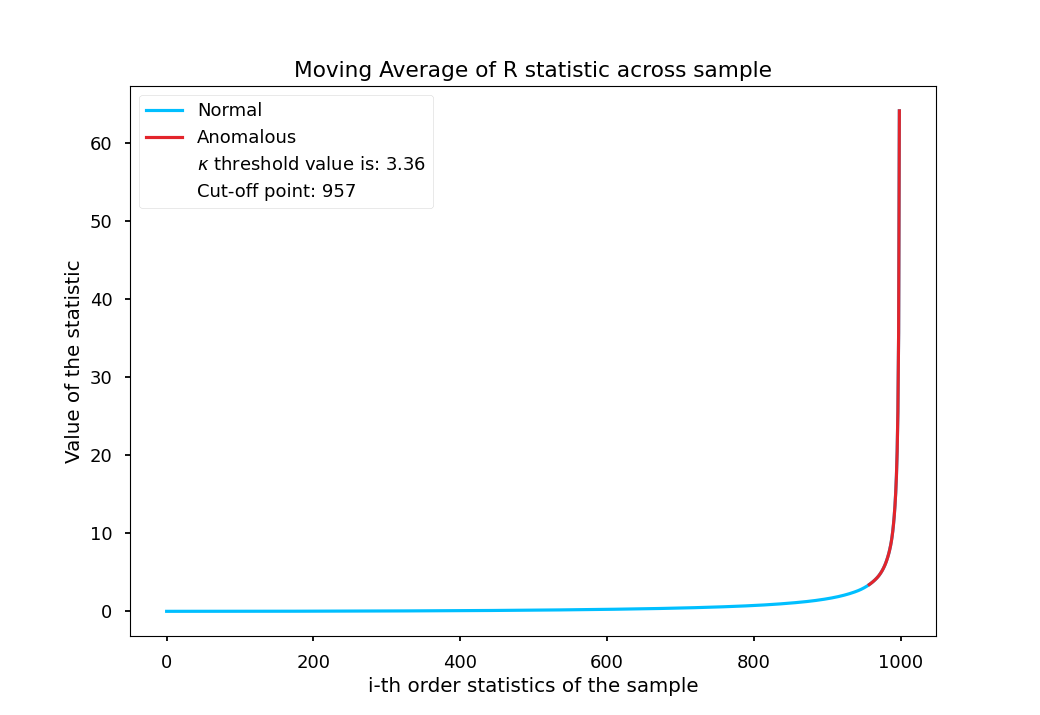} \\
    \end{tabular}
\end{minipage}
\caption{Simulation studies in figure \ref{plot:distributions} revisited with $\kappa$-threshold chosen automatically.}
\label{plot:distributions3}
\end{figure}

In the Figure \ref{plot:distributions3} we revisit our simulation studies and show the simulation studies and show the chosen cut-off points using the kneedle algorithm, choosing the sensitivity parameter as $5.0$.

\subsection{Application II: Distinguishing two Pareto tails}\label{subsection:tails}
Now that we have a method for selecting a $\kappa$ value, we conduct further simulation tests based on distinguishing two Pareto tails in order to show the efficacy of our statistic. Our goal is to find a threshold $\kappa$ beyond which $ >.90$ of the observations belong to the "heavier" tail. We set up our experiment as follows. For given two tail indices $\alpha_1 < \alpha_2$, we first determine a $\kappa$-threshold value for the $\alpha_1$ indexed Pareto distribution. Then we sample a $N = 1000000$ observations from both of the distributions and calculate our statistic across the entire sample. We report the percentage of observations from $\alpha_2$ in the samples above the predetermined threshold. We repeat this experiment a $1000$ times in order to build a confidence interval on our results.

A key problem in our experiments is the selection $\alpha_1$ and $\alpha_2$ values, it has been established that as the two indices get closer it becomes numerically difficult to differentiate between the tails, even if one of the tails are outside of the Levy-stable regime \cite{Weron_2001}. We start our experiments with indices $\alpha_1 = 1.5$ and $\alpha_2 = 2.5$ and move $\alpha_2$ closer every iteration.

The results of the experiments are given in the table \ref{table:pareto_sim} below. When the tail indices are far apart our statistic produces an ideal change point for differentiating between the two tails. It is clear that as the two tail indices get closer our statistic becomes unable to distinguish the difference between the two tails. However, it remains consistent with variance $\sim 10^{-5}$ across all samples.

\begin{table}[H]
\begin{center}
    \begin{tabular}{||c c c c c||} 
    \hline
    $\alpha_1$ & $\alpha_2$ & $\kappa$-threshold & Expected percentage of $\alpha_2$ above $\kappa$ & Variance \\ [0.5ex] 
    \hline\hline
    1.5 & 2.5 & 2.745 & 0.95 & $1.88\times10^{-5}$ \\
    \hline\hline
    1.5 & 2.3 & 2.745 & 0.924 & $4.06\times10^{-5}$ \\ 
    \hline
    1.5 & 2.1 & 2.745 & 0.85 & $1\times10^{-4}$ \\
    \hline\hline
    1.5 & 1.9 & 0.78 & 0.95 & $7.37\times10^{-5}$ \\ 
    \hline
    1.5 & 1.7 & 2.745 & 0.65 & $3.74\times 10^{-5}$ \\
    \hline
    \end{tabular}
\end{center}
\caption{Simulation results of two Pareto tails.}
\label{table:pareto_sim}
\end{table}

\section{Results and Discussions}
In this work we stray away from an all-encompassing definition of an outlier, rather focus on a definition for the one dimensional case in terms of order statistics. In a sense our definition tries to approach the problem from the point of view of investigating data points "that arouse suspicions that it was generated by a different mechanism" \cite{Hawkins_1980}.

There are a few good reasons motivating our decision. Similar to the case in extreme value theory, it is difficult to order random vectors and the limit cases are not as intuitive \cite{DeHaan_2006}. Furthermore, any multivariate definition must take into consideration cases where two or more random variables are not independent. Since the requirement of independence is too stringent for applied studies as conditional outliers are all too common an occurrence in real life \cite{Chandola_2009}.

A particular strong point of our method is the ability to select cut-off points for discrete set of points without the need of a priori information. As a result, our definition is innately compatible with any definition which produces outlier scores. For the multivariate case, since much of the literature is applied, we recommend the reader to first use the method which suits them best. Our definition, together with the $\kappa$-threshold selection, can be used afterwards to select a natural cut-off point.

We calculate the case $S_{m-1}/T_{n-m}$ and approximate for the defined $R$-statistic $S_{m}/T_{n-m}$. From the approximation, we can see that concentration of $R$-statistic depends on the tail of the random variable. Our simulation studies in \ref{section:experiments} also confirmatory. In particular, in \ref{plot:distributions2} it is considerably more difficult to find a cut-off point for the half-normal and exponential distributions compared to the identified outliers model. Nevertheless, we see in section \ref{subsection:tails} that even in the cases where the tail index is not in L\'{e}vy stable region, our statistic still produces results with very small variance with varying degrees of success.


For future work, we may choose the possible cut-off point by using the $R$-statistic on moving blocks first, then selecting the candidate blocks based on results. Finally, we can find the threshold value by looking at the block with the most dramatic increase. Identify potential blocks for the cut-off point is also helpful for the cases when a practitioner wishes to pick the value the $\kappa$-threshold by hand.

\clearpage
\bibliography{IEEEabrv,article}
\end{document}